\documentclass{aa}  
\usepackage{graphicx}
\usepackage{txfonts}
\usepackage{natbib}
\bibpunct{(}{)}{;}{a}{}{,} % to follow the A&A style
\newcommand{\mincir}{\raise -2.truept\hbox{\rlap{\hbox{$\sim$}}\raise5.truept
\hbox{$<$}\ }}
\newcommand{\magcir}{\raise -2.truept\hbox{\rlap{\hbox{$\sim$}}\raise5.truept
\hbox{$>$}\ }}
\newcommand{\siml}{\raise -2.truept\hbox{\rlap{\hbox{$\sim$}}\raise5.truept
\hbox{$<$}\ }}
\newcommand{\simg}{\raise -2.truept\hbox{\rlap{\hbox{$\sim$}}\raise5.truept
\hbox{$>$}\ }}
\newcommand{\be}{\begin{equation}}
\newcommand{\ee}{\end{equation}}
\newcommand{\ba}{\begin{eqnarray}}
\newcommand{\ea}{\end{eqnarray}}
\newcommand {\kpc} {kpc $\;$}

\newcommand {\mpc} {Mpc $\;$}

\newcommand {\h} {Mpc$\;$}
\newcommand {\hh} {Mpc}

\newcommand {\ks} {km~s$^{-1} \;$}
\newcommand {\kss} {km~s$^{-1}$}

\newcommand {\mqua} {$\times 10^{14}\;M_{\odot} \;$}
\newcommand {\mquaa} {$\times 10^{14}\;M_{\odot}$}

\newcommand {\ml} {$M_{\odot}/L_{\odot} \;$}
\newcommand {\mll} {$M_{\odot}/L_{\odot}$}

\newcommand{\degree}{\ensuremath{\mathrm{^\circ}}}
\newcommand{\arcm}{\ensuremath{\mathrm{^\prime}\;}}
\newcommand{\arcs}{\ensuremath{\arcmm\hskip -0.1em\arcmm \;}}
\newcommand{\arcmm}{\ensuremath{\mathrm{^\prime}}}

\newcommand{\dotarcs}{\,\rlap{\hbox{$\mathrm{^\prime\hskip-0.1em^\prime}$}}{\hbox{$.$}}\,}

\newcommand{\dotsec}{\,\rlap{\hbox{$\mathrm{^s}$}}{\hbox{$.$}}\,}
\begin{document}

\title{Accretion processes in the galaxy cluster Hydra~A/Abell~780 \thanks{Full Table 1 is only available in electronic form
at the CDS via anonymous ftp to cdsarc.u-strasbg.fr (130.79.128.5)
or via http://cdsweb.u-strasbg.fr/cgi-bin/qcat?J/A+A/}}

%   \subtitle{I. Overviewing the $\kappa$-mechanism}
%
%   \author{G. Wuchterl
%          \inst{1}
%          \and
%          C. Ptolemy\inst{2}\fnmsep\thanks{Just to show the usage
%          of the elements in the author field}
%          }

\author{M. Girardi\inst{1,2},
W. Boschin\inst{3,2,4,5},
M. Nonino\inst{2},
C. Innocentin\inst{1},
S. De Grandi\inst{6}}

%  \offprints{M. Girardi, \email{marisa.girardi@inaf.it}}

  \institute{Dipartimento di Fisica dell'Universit\`a degli Studi di Trieste -
Sezione di Astronomia, via Tiepolo 11, I-34143 Trieste, Italy \email{marisa.girardi@inaf.it}
\and INAF - Osservatorio Astronomico di Trieste, via Tiepolo 11,
I-34143 Trieste, Italy
\and Fundaci\'on Galileo Galilei - INAF (Telescopio Nazionale
  Galileo), Rambla Jos\'e Ana Fern\'andez Perez 7, E-38712 Bre\~na
  Baja (La Palma), Canary Islands, Spain
\and Instituto de Astrof\'{\i}sica de Canarias, C/V\'{\i}a L\'actea
s/n, E-38205 La Laguna (Tenerife), Canary Islands, Spain
\and Departamento de Astrof\'{\i}sica, Univ. de La Laguna, Av. del
Astrof\'{\i}sico Francisco S\'anchez s/n, E-38205 La Laguna
(Tenerife), Spain
\and INAF - Osservatorio Astronomico di Brera, via E. Bianchi 46, I-23807 Merate, Italy
  }

  \date{Received  / Accepted }

  \abstract {Clusters of galaxies evolve and accrete mass, mostly from
    small galaxy systems.}{Our aim is to study the velocity field of
    the galaxy cluster Abell~780, which is known for the powerful
    radio source Hydra~A at its center and where a spectacular X-ray
    tail associated with the galaxy LEDA~87445 has been discovered.}
            {Our analysis is based on the new spectroscopic data for
              hundreds of galaxies obtained with the Italian
              Telescopio Nazionale {\em Galileo} and the Very Large
              Telescope. We have constructed a redshift catalog of 623
              galaxies and selected a sample of 126 cluster
              members. We analyze the internal structure of the
              cluster using a number of techniques.}  {We estimate the
              mean redshift $z = 0.0545$, the line-of-sight velocity
              dispersion $\sigma_{\rm V}\sim 800$ \kss, and the
              dynamical mass $M_{200}\sim 5.4$ \mquaa.  The global
              properties of Abell~780 are typical of relaxed clusters.
              On a smaller scale, we can detect the presence of a
              galaxy group associated with LEDA~87445 in projected
              phase space. The mean velocity and position of the
              center of the group agree well with the velocity and
              position of LEDA~87445. We estimate the following
              parameters of the collision. The group is characterized
              by a higher velocity relative to the main system. It is
              infalling at a rest frame velocity of $V_{\rm rf}\sim
              +870$ \kss and lies at a projected distance $D\sim 1.1$
              Mpc to the south, slightly southeast of the cluster
              center. The mass ratio between the group and the cluster
              is $\sim$1:5. We also find evidence of an
                  asymmetry in the velocity distribution of galaxies
                  in the inner cluster region, which might be related
                  to a small low-velocity group detected as a
                  substructure at $V_{\rm {rf}}\sim -750$ \kss.}
            {We conclude that A780, although dynamically relaxed at
              first sight, contains small substructures that may have
              some impact on the energetics of the core region.}


\titlerunning{Accretion in Abell~780} 
\authorrunning{Girardi et al.} 

\maketitle

\section{Introduction}
\label{intro}

It is well known that galaxy clusters are not simple relaxed systems,
but that they evolve and increase in mass through a hierarchical
merging process from poor groups to rich clusters. Large mergers
involving two galaxy systems of comparable mass are the most energetic
events in the Universe since the Big Bang. However, as numerical
simulations show, accretion of smaller systems at the galaxy or
group scale is the main channel of cluster growth (e.g.,
\citealt{berrier2009}; \citealt{mcgee2009};
\citealt{benavides2020}). Observational evidence for cluster accretion
in the past comes from the presence of cluster substructures detected
in the distributions of member galaxies, gas, and mass as determined
by optical, X-ray, and gravitational lensing data
(\citealt{feretti2002} and \citealt{molnar2015} for
reviews). Detecting and studying the accretion of small galaxy systems
is more difficult than that of mergers of the same size.  Indeed, this
requires the use of several hundred redshifts per cluster and/or the
isophotes of X-ray brightness residuals (e.g., \citealt{adami2005} in
Coma cluster; \citealt{girardi2015} for the CLASH-VLT survey;
\citealt{lisker2018} in Virgo cluster).

The detection of accreted small groups, and in particular their member
galaxies, is a fundamental step in understanding the effects of the
cluster environment on galaxies. Indeed, one of the key issues is to
understand the dicothomy of galaxy types between the cluster and the
field environment, both in terms of the mechanism that stalls star
formation in galaxies (e.g., \citealt{peng2015}) and whether this is
related to the environment of the cluster (e.g., \citealt{treu2003})
or with the earlier preprocessing in groups (e.g.,
\citealt{vijayaraghavan2013}).

Observations of X-ray bright head-tail downstreams are generally
interpreted as direct evidence for the ongoing process of cluster
accretion, in which the gaseous component of the infalling galaxies
(or galaxy groups) is stripped by the intracluster medium (ICM). The
observation of these features is quite rare, probably due to the high
quality of the X-ray data required. As a concrete example, one can
cite M86 in Virgo (\citealt{forman1979}), NGC 1404 in Fornax
(\citealt{jones1997}), A2142 (\citealt{eckert2014}), Hydra~A
(\citealt{degrandi2016}), and Abell~85 (\citealt{ichinohe2015}).  In
practice, these X-ray bright tail streams could serve as reliable
signposts for infalling galaxies or groups into the cluster
environment.

\citet{degrandi2016} report the discovery of a structure with a
spectacular bright X-ray tail in the outskirts of the
\object{Abell~780} cluster (hereafter A780) using deep wide-field
observations with \emph{XMM-Newton} and \emph{Suzaku}. They suggest
that this feature is related to an accreting galaxy group that is
clustered around the bright galaxy LEDA~87445, $\sim 17^{\prime}$
($1.1$ Mpc) south of the cluster center, with a wake of stripped gas
extending behind the group over a projected scale of 760 kpc. The
temperature of the gas along the wake is constant at $kT_{\rm X}\sim
1.3$ keV, which is about a factor of two below the temperature of the
surrounding ICM.

A780 is a modest cluster, characterized by low richness (Abell richness
class $R=0$, \citealt{abell1989}), X-ray luminosity ($L_{\rm
  {X,bol}}\sim 2\times10^{44}$ erg s$^{-1}$, \citealt{zhang2011}), and
temperature ($kT_{\rm X}\sim 3-3.5$ keV, \citealt{sato2012},
\citealt{degrandi2016}). These X-ray values are typical for small 
clusters as shown by the relation $L_{\rm X}-T_{\rm X}$ (see Fig.~13 of
\citealt{osmond2004}).

The A780 cluster shows typical properties of a dynamically relaxed
galaxy system. \emph{Chandra} and \emph{XMM-Newton} observations
indicate the presence of a cool core (\citealt{mcnamara2000}; see also
Fig.~2 of \citealt{simionescu2008}).  \emph{XMM-Newton} observations
show that the X-ray morphology of A780 is smooth within 1 Mpc (see,
e.g., Fig.~1 of \citealt{degrandi2016}).  As for the galaxy component,
no difference in the luminosity function of galaxies was detected in
different regions within 500 kpc (\citealt{durret2009}).

Instead, A780 shows many small-scale substructures related to the
AGN-ICM interaction in the center of the cluster. In fact, A780 is
better known for the powerful radio source, \object{Hydra A} (3C~218),
emanating from its central, brightest galaxy (\citealt{matthews1964}).
The radio source has a complex morphology that extends nearly $\sim$
8\arcm ($500$ kpc) in the north-south direction, slightly NNE-SSW
(\citealt{lane2004}), while active jets have been detected in the
central $\sim50$ kpc region (\citealt{taylor1990}). \emph{Chandra}
X-ray observations have revealed a multicavity system, a large-scale
shock, an X-ray bright filament near the cluster center, and a spiral
like structure in the ICM (\citealt{mcnamara2000},
\citealt{nulsen2005}; \citealt{wise2007};
\citealt{lagana2010}). According to general consensus, the giant radio
lobes, triggered by AGN jets interact with the surrounding ICM and
regulate the cooling flow in Hydra~A.

There are few spectroscopic data in the literature for galaxies in the
field of A780. \citet{durret2009} analyzed the region of 5 deg radius
($\sim$ 19 Mpc) around the center of the cluster using galaxies with
available redshift in the NASA/IPAC Extragalactic Database (NED). No
obvious large-scale structure was found within 5 Mpc of the cluster,
and the cluster is likely detected as a bound structure of 14
galaxies. To date, the only homogeneous, one-source redshift catalog
is that of \citet{smith2004}, who list data for 41 galaxies and one
star as part of their wide-field imaging/spectroscopic study of 93
clusters (NOAO Fundamental Plane Survey). We have previously analyzed
these data and extracted a subset of 33 fiducial member galaxies
(\citealt{degrandi2016}, see Appendix A). In that work we have
presented the first evidence of a substructure in the galaxy
distribution around LEDA~87445, but more data are needed to determine
the kinematic and dynamical properties of the group and the parameters
of the collision.

In this context, we have carried out an intensive observational
campaign to obtain new spectroscopic data with both the Italian
Telescopio Nazionale {\em Galileo} (TNG) and the Very Large Telescope
(VLT). This paper is devoted to the presentation of our results on the
velocity field of A780. The paper is organized as follows. We describe
the optical observations and present our spectroscopic data catalog in
Sects.~2 and 3. We select the members of the cluster, estimate the
global properties, and analyze the substructure in Sects.~4 and
5. Section~6 is devoted to the interpretation and discussion of our
results. In Sect.~7, we give a brief summary and derive our
conclusions.  In this work we use $H_0=70$ km s$^{-1}$ Mpc$^{-1}$ in a
flat cosmology with $\Omega_0=0.3$ and $\Omega_{\Lambda}=0.7$. In the
assumed cosmology, 1\arcm corresponds to $\sim 63.5$ \kpc at the
cluster redshift.  Recall that the velocities we derive for the
galaxies are line-of-sight velocities determined from the redshift,
$V=cz$. Unless otherwise stated, we report errors with a confidence
level (c.l.) of 68\%.

\section{New spectroscopic data}
\label{data}

Multi-object spectroscopic observations of A780 were performed  at
the TNG in February 2016 (program AOT32-TAC25, PI: M. Girardi). We
used the instrument DOLoRes in MOS mode with the LR-B
Grism\footnote{http://www.tng.iac.es/instruments/lrs}. In total, we
observed five  MOS masks for a total of 194 slits. The total exposure time
was 3600 s for each of the five masks.

Spectral reduction and redshift estimation were performed using
standard IRAF\footnote{IRAF is distributed by the National Optical
  Astronomy Observatories, which are operated by the Association of
  Universities for Research in Astronomy, Inc., under a cooperative
  agreement with the National Science Foundation.} tasks and the
cross-correlation technique (\citealt{tonry1979}). The line-of-sight
velocity errors are a result of the cross-correlation technique. In 43
cases, the redshift was estimated by measuring the wavelength position
of emission lines in the spectra. Our spectroscopic catalog contains
165 galaxies. For five galaxies there are double redshift
determinations measured in different masks.

We have checked the consistency of redshift measurements in different
masks using the following procedure (see, e.g.,
\citealt{barrena2007fit}; hereafter FIT-ZZ procedure). For the five
galaxies with double redshift determination, we fitted the first
measurement against the second by using a straight line and accounting
for errors in both coordinates (\citealt{press1992}). The fitted line
agrees with the one-to-one relation, but the value of the
$\chi^2$-probability indicates that the errors are too small to
explain the observed scatter. Only when the nominal errors of the
redshifts are multiplied by a factor of two can the observed scatter be
justified. We have corrected the error estimates accordingly.  For the
galaxies with two redshift measurements, we combined the two redshift
determinations using weighted averages and errors. Our final 
spectroscopic TNG catalog lists 160 galaxies brighter than $r$=21 mag,
most brighter than $r$=20 mag. The median value of the $cz$ errors is
98 \kss.

Multi-object spectroscopic observations of Hydra~A were performed at
the VLT in January, February, and March 2017 (Program 098.A-0807(A),
P.I.: M. Girardi). We used the instrument VIMOS in MOS mode with the
LR-blue grism. In total, we observed three MOS mask pointings for a
total of 1407 slitlets. The total exposure time was six hours,
distributed over three pointings. Spectral reduction and calculation of
radial velocities were performed using IRAF as described above.  We
succeeded in obtaining a reliable determination of the redshift for
478 galaxies. The success rate of $30\%$ was mainly due to the poor
weather conditions. We obtained duplicate determinations for 20
galaxies measured in different masks, and we followed the FIT-ZZ
procedure described above to verify the consistency of measurements.
This analysis shows that the errors are underestimated by a factor 2.5
and we corrected them accordingly.  For the galaxies with two redshift
measurements, we combined the duplicate determinations as above.  Our
final spectroscopic VLT catalog lists 458 galaxies down to $r\sim$24
mag, about three magnitudes fainter than the data from TNG.  The
median value of the $cz$ errors is 140 \kss.

\section{Spectroscopic catalog}
\label{cata}

In creating a unique catalog, we also took into account the redshift
data provided by Smith et al. (\citeyear{smith2004}, hereafter S04
catalog, 41 galaxies), which have already been analyzed in our
previous study (\citealt{degrandi2016}). We verified the consistency
of the S04 data with our new TNG data by applying the FIT-ZZ procedure
to the 13 objects listed in both catalogs. This analysis shows that
the S04 errors are underestimated by a factor of two and we corrected them
accordingly. For the galaxies listed in both catalogs, we combined the
duplicate redshift determinations as above.
We obtained the combined catalog TNG+S04 and compared it
with the VLT catalog. We verified their consistency by
applying the FIT-ZZ procedure to 23 common objects and combined the
data to obtain a unique catalog.  The final spectroscopic catalog
consists of 623 galaxies, mostly from VLT observations, covering
mainly the cluster core and the region around LEDA~87445.

We have also used public photometric data from the CFHT archive obtained with
the Megaprime/Megacam camera at the Canada-France-Hawaii Telescope.
From the CADC Megapipe archive (\citealt{gwyn2009}) we retrieved the catalogs
for the images in Megacam $g$ and $r$ bands and corrected the
magnitudes for Galactic extinction. The total area covered by the
images is 1.05$\times$1.16 deg$^2$. The estimated limiting magnitudes
(at the $5\sigma$ c.l.) are $g$ = 25.5 mag and $r$ = 24.5 mag.

Table~\ref{catalogA780}, available at CDS, lists the velocity catalog
for the member galaxies, as defined in Sect.~\ref{memb}:
identification number of each galaxy, ID (Col.~1); right ascension and
declination, $\alpha$ and $\delta$ (J2000, Col.~2); $r$ magnitude
(Col.~3); heliocentric radial velocities, $V=cz$ (Col.~4), with
errors, $\Delta V$ (Col.~5); source of spectroscopic data (V: VLT, T:
TNG, L: S04).

\begin{figure*}[!ht]
%\begin{figure*}
\centering 
\includegraphics[width=18cm]{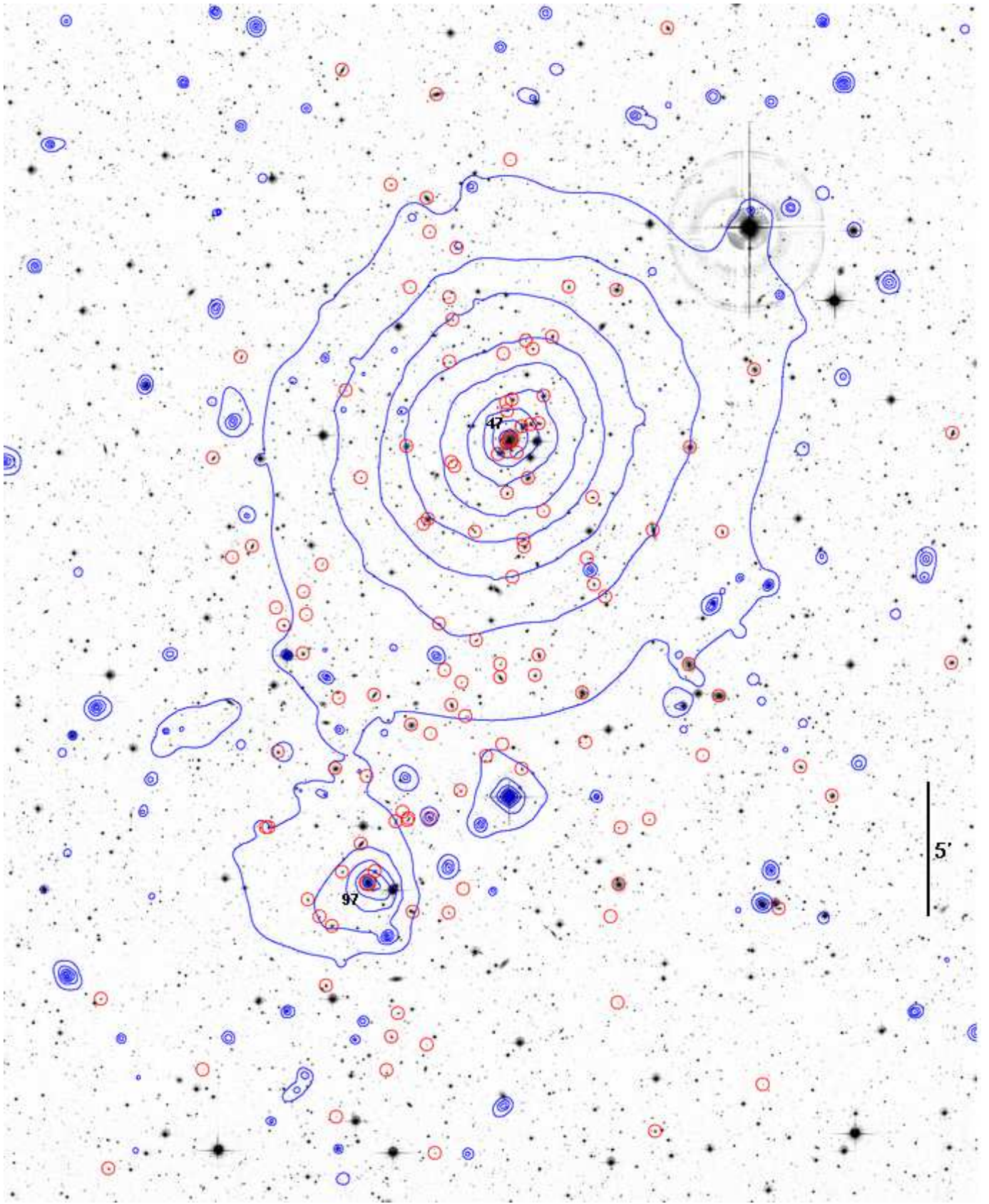}
\caption{CFHT/Megacam $r$-band image (north top and east left) of
  the galaxy cluster A780 with the contour levels of the \emph{XMM-Newton} X-ray
  image superimposed (photons in the energy range 0.7--1.2 keV,
  \citealt{degrandi2016}). Cluster members, as defined in
  Sect.~\ref{memb}, are highlighted by red circles (see
  Table~\ref{catalogA780}). The labels denote the galaxies discussed in the
  text, ID.~47 (the BCG where the Hydra~A radio source is located) and
  ID.~97 (LEDA~87445).  }
\label{figottico}
\end{figure*}

Figure~\ref{figottico} shows the cluster region sampled by our
redshift catalog, with the \emph{XMM-Newton} X-ray contours
superimposed.  Our new spectroscopic observations are mainly focused on the
south-southeastern part of A780, allowing us to analyze the cluster
galaxies from the cluster center out to LEDA~87445.

\begin{table}
        \caption[]{Radial velocities of 126 member galaxies of A780}
        \label{catalogA780}
              $$ 
        % \begin{array}{p{0.5\linewidth}l}
            \begin{array}{r c c r r c}
            \hline
            \noalign{\smallskip}
            \hline
            \noalign{\smallskip}

\mathrm{ID} & \alpha,\delta\,(\mathrm{J}2000) & r\, &V\,& \Delta V& \mathrm{Source}\\
 &                              & &\mathrm{\,(\,km}&\mathrm{s^{-1}\,)}& \\
            \hline
            \noalign{\smallskip}  

1    & 9\ 16\ 58.13,-12\ 05\ 28.6 &   16.50 & 16338&   50& \mathrm{L} \\  
2    & 9\ 16\ 58.23,-12\ 14\ 06.4 &   16.55 & 16200&   58& \mathrm{L} \\  
3    & 9\ 17\ 16.46,-12\ 19\ 07.4 &   16.73 & 16988&   72& \mathrm{L} \\  
4    & 9\ 17\ 21.29,-12\ 18\ 02.0 &   16.92 & 16583&   56& \mathrm{L} \\  
5    & 9\ 17\ 24.58,-12\ 23\ 21.7 &   17.35 & 14401&  130& \mathrm{L} \\
6    & 9\ 17\ 27.02,-12\ 29\ 59.5 &   20.63 & 16534&  210& \mathrm{V} \\

            \noalign{\smallskip}			          
            \hline					    
            \end{array}					 
            $$
            \tablefoot{Full table is available at CDS.}
\end{table}

%%%%

\begin{figure}
\centering
\includegraphics[width=8.5cm]{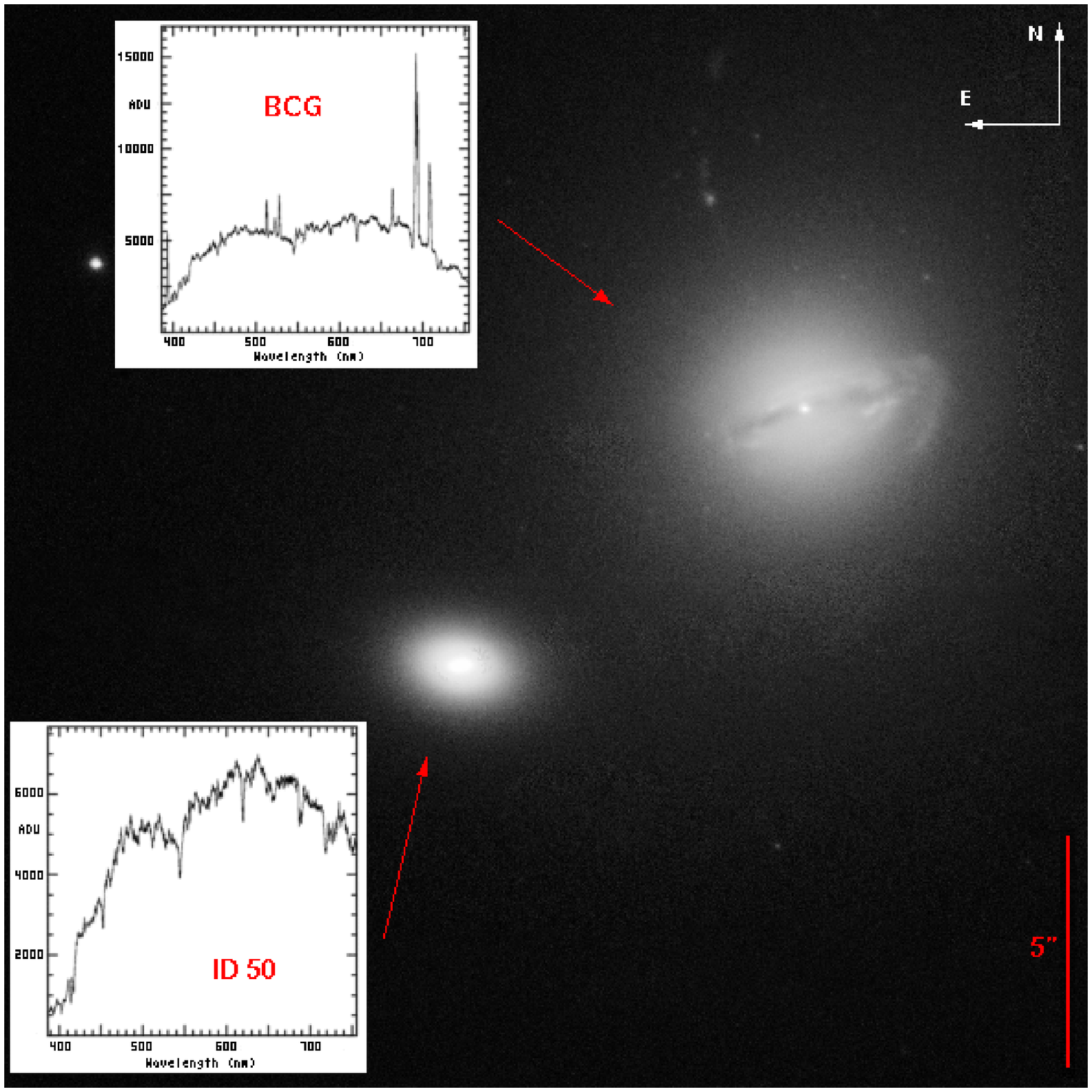}
  \caption{NIR diffraction-limited image of the BCG and ID~50 taken
    with the MCAO system of the Gemini South telescope (adapted from
    Fig.~5 of \citealt{neichel2014}). The insets show
    the wavelength-calibrated TNG spectra of BCG and its close projected
    companion.}
\label{figneichel}
\end{figure}

The cluster galaxy population is dominated by the brightest cluster
galaxy (BCG) ID~47, which is located in the center of the X-ray
emission. The BCG was described as a cD galaxy with an unresolved
binary nucleus (\citealt{dewhirst1959,matthews1964}). The secondary
nucleus was found to be blueshifted with respect to the primary
nucleus. Indeed, a diffraction-limited $K_{\rm s}$-band image taken
with the MCAO system of Gemini South (\citealt{neichel2014}) shows
that there is a close companion galaxy to the southeast of the BCG
(see also our Fig.~\ref{figneichel}). In our catalog, the companion
galaxy is identified as ID~50 and is blueshifted of $\sim 900$ \ks
with respect to the BCG. Both galaxies lie within the velocity
distribution of member galaxies and are separated by a value slightly
larger than the velocity dispersion (see below,
Sect.~\ref{memb}). This confirms that ID~50 is an independent cluster
galaxy projected onto the cluster center, very close to the BCG.
Another prominent galaxy is LEDA~87445 (ID.~97), the galaxy associated
with the bright X-ray tail (\citealt{degrandi2016}).

\section{Member selection and global properties}
\label{memb}

To select cluster members among the 623 galaxies in our spectroscopic
catalog, we used the two-step method known as the ``Peak+Gap'' (P+G),
which was previously applied by \citet{girardi2015}.  The first step
is the application of the 1D adaptive-kernel DEDICA method
(\citealt{pisani1993} and \citealt{pisani1996}; see also
\citealt{girardi1996}).  This method detects A780 as a peak at
$z\sim0.0543$ populated by 131 galaxies. All but one the non-members
are background galaxies.

\begin{figure}
\centering
\resizebox{\hsize}{!}{\includegraphics{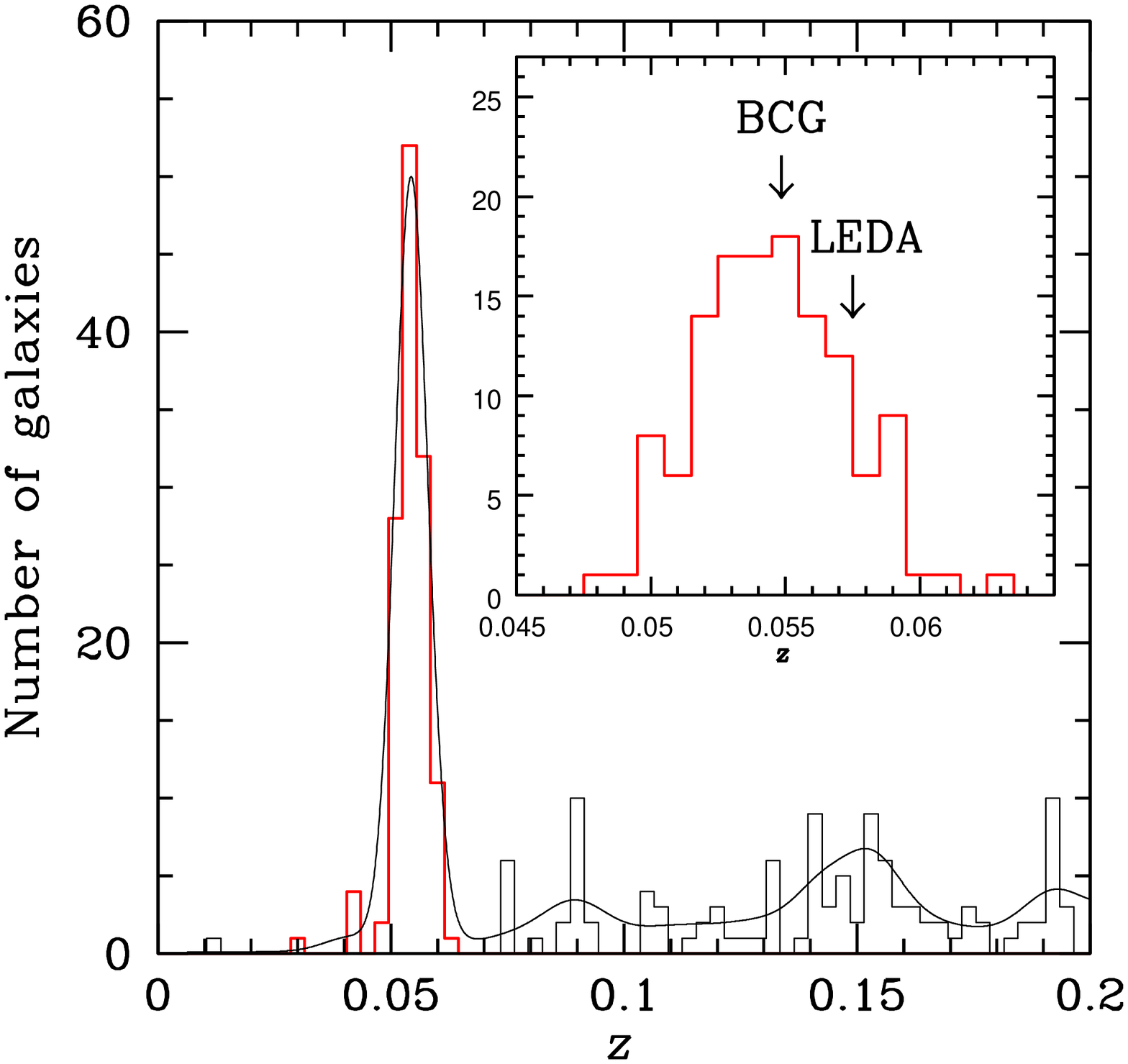}}
\caption
{Distribution of galaxy redshifts within $z=0.2$. The histogram refers
  to all galaxies with the spectroscopic redshift in the region of
  A780.  The histogram with the thick, red line refers to the 131
  galaxies assigned to the A780 peak by the 1D DEDICA reconstruction
  method (faint line).  The inset figure shows the 126 member galaxies
  with the redshifts of the BCG and LEDA~87445 indicated.}
\label{fighisto}
\end{figure}

In a second step, we combine the spatial and velocity information by
applying the ``shifting gapper'' procedure (\citealt{fadda1996,girardi1996}). Of the galaxies that lie within an annulus around from the center of cluster, this
procedure rejects those that are too far away in velocity from
the main body of galaxies, i.e. farther away than a fixed velocity gap. The
position of the annulus is shifted at increasing distances from
the center of the cluster. The procedure is repeated until the number of
cluster members converges to a stable value.  Following
\citet{fadda1996}, we use a gap of $1000$ \ks in the cluster
rest-frame, and a annulus size of 0.6 \h  or larger at least to
include 15 galaxies.  In fixing the cluster center, we consider the
position in R.A. and Dec. of the BCG
[R.A.=$09^{\mathrm{h}}18^{\mathrm{m}}05\dotsec66$, Dec.=$+12\degree
  05\arcmm 43\dotarcs7$ (J2000.0)]. This procedure confirms 126
fiducial cluster members in the range $14401 \leq{\rm V} \leq
18745$ \kss, i.e. $0.048036 \leq {\rm z} \leq 0.062527$ 
(see Fig.~\ref{fighisto}).

\begin{figure}
\centering
\resizebox{\hsize}{!}{\includegraphics{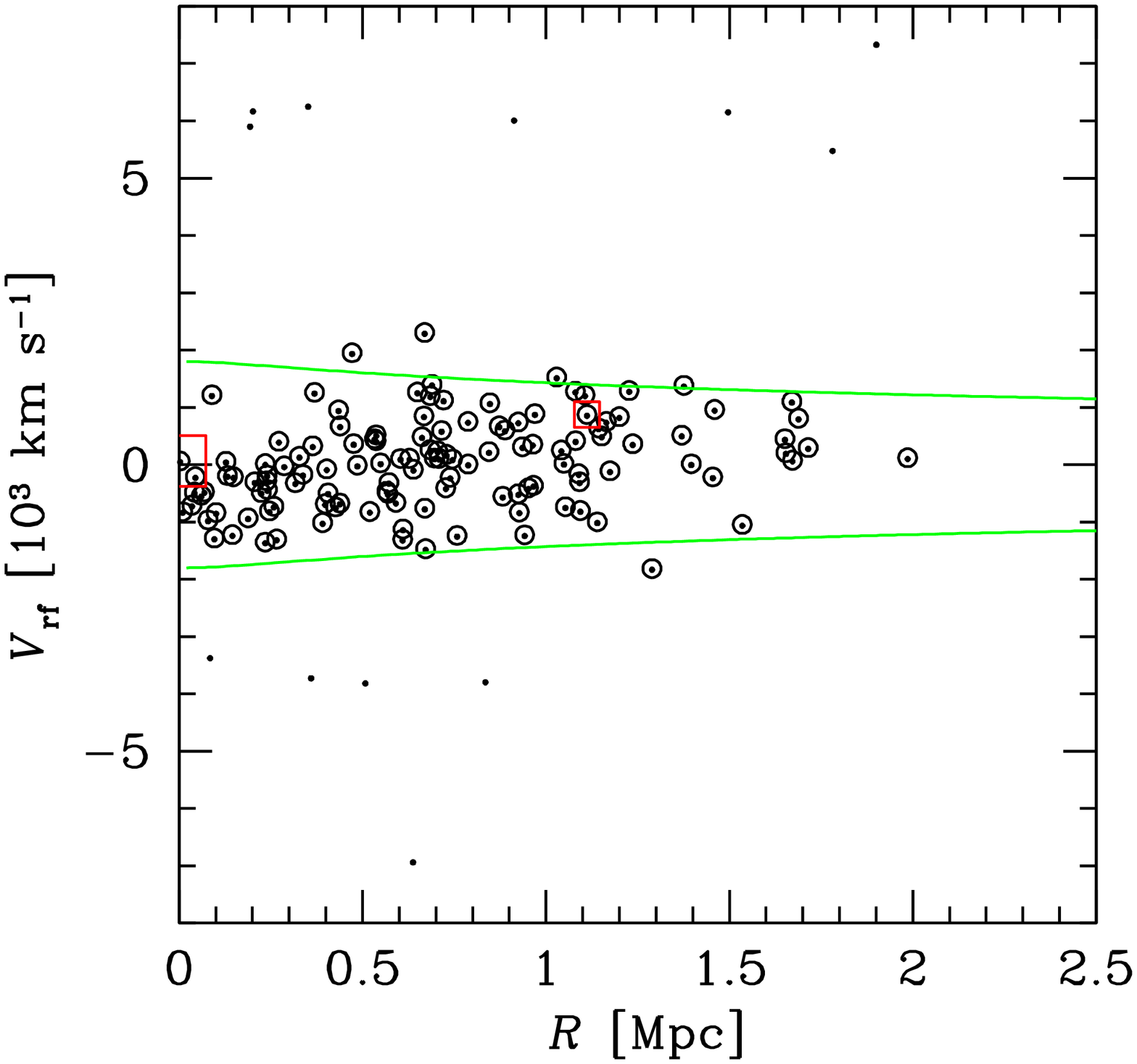}}
\caption
{Rest-frame velocity $V_{\rm rf}=(V-<V>)/(1+z)$ versus projected
  cluster-centric distance $R$ for galaxies with redshifts in the
range   $\pm8000$ \ks (black dots). Circles indicate
  the 126 members of the cluster. The large and
  small red squares refer to the BCG and LEDA~87445, respectively. The green
  curves contain the region where $|V_{\rm rf}|$ is smaller than the
  escape velocity (see text).  }
\label{figvd}
\end{figure}

Although A780 is a modest cluster, it is well isolated in 
projected phase space, as shown in Fig.~\ref{figvd}.  Just to
highlight the region of cluster members, we also plot the
escape velocity curves obtained with the mass estimate calculated below
and assuming an NFW mass density profile following 
the recipe of \citet{denhartog1996}.

The analysis of the velocity distribution of the 126 cluster members
was performed using the biweight estimators for location and scale
(ROSTAT software, \citealt{beers1990}). Our measurement of the mean 
redshift of the cluster is $\left<z\right>=0.0545\pm0.0002$ (i.e.,
$\left<V\right>=16313\pm71$ \kss).  We estimate the velocity
dispersion, $\sigma_{\rm V}$, by applying the cosmological correction and
the standard correction for velocity errors (\citealt{danese1980}). We obtain
$\sigma_{\rm V}=795_{-53}^{+44}$ \kss, where the errors are estimated by a
bootstrap technique.

Using the fitted relation between the mass\footnote{We refer to $R_{\Delta}$
  as the radius of a sphere within which the mean mass density is
  $\Delta$ times the critical density at the redshift of the galaxy
  system. $M_{\Delta}$ is the mass contained in $R_{\Delta}$.}
$M_{200}$ and the velocity dispersion in simulated clusters 
(Eq.~1 of \citet{munari2013} we derive $M_{200}=5.4\pm1.5$ \mqua for the
mass contained in $R_{200}=1.65\pm0.10$ Mpc. The uncertainties of
6\% and 28\% for $R_{200}$ and $M_{200}$ are calculated using the
error propagation for  $\sigma_{\rm V}$ ($R_{200}\propto \sigma_{\rm
  V}$ and $M_{200}\propto \sigma_{\rm V}^3$) and an additional 
uncertainty of $10\%$ for the mass due to the scatter around the relation of
\citet{munari2013}. Properties of the cluster are shown in
Tab.~\ref{tabv}.

\begin{table*}
        \caption[]{Global properties of A780}
         \label{tabv}
            $$
         \begin{array}{c c c c c c}
            \hline
            \noalign{\smallskip}
            \hline
            \noalign{\smallskip}

            N_{\rm gal} &^{\mathrm{a}}\alpha({\rm J}2000),\,\delta({\rm J}2000)&
            \mathrm{<V>}&\sigma_{\rm V}&R_{200}&M_{200}\\
            &\mathrm{h:m:s,\degree:\arcmm:\arcs}&
            \mathrm{km\ s^{-1}}&\mathrm{km\ s^{-1}}&{\rm Mpc}&10^{14}M_{\odot}\\

            \hline
            \noalign{\smallskip}

         126&09\ 18\ 05.66,-12\ 05\ 43.8&16313\pm71&795_{-53}^{+44} &1.65\pm0.10&5.4\pm1.5\\
            \noalign{\smallskip}
            \hline
         \end{array}
         $$
         
\begin{list}{}{}  
\item[$^{\mathrm{a}}$] As a center, we list the position of the BCG. 
\end{list}
         \end{table*}

\begin{figure}
\centering
\resizebox{\hsize}{!}{\includegraphics{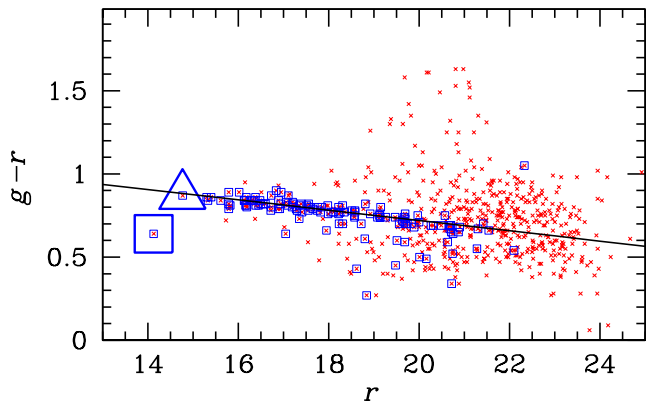}}
\caption
    {$g$$-$$r$ vs. $r$ diagram (aperture color  
        vs. Kron-like magnitude). All galaxies with
      available spectroscopy and member galaxies are indicated by red
      crosses and blue squares, respectively.  The large, blue square and
      triangle indicate the BCG and LEDA~87445, respectively.  The
      black line shows the color-magnitude relation obtained
      for the member galaxies.  }
\label{figcm}
\end{figure}

Figure~\ref{figcm} shows the distribution of the member galaxies in
the ($g$$-$$r$ vs. $r$) color-magnitude diagram, compared to all
galaxies with redshift measurements. The color-magnitude relation,
indicating the location of the early-type galaxies, can be seen out to
faint magnitudes of $r\sim$21 mag. Following
\citet{boschin2012twosigma} we use a $2\sigma$ rejection procedure to
fit $g$$-$$r$=1.34$-$0.03$\times$ $r$.

\section{Substructure analysis}
\label{sub}

We analyzed the presence of substructures using the velocity
distribution of galaxies, their projected positions on the sky,
and the combination of these two pieces of information (1D, 2D, and 3D tests,
respectively).

Using a number of indicators such as kurtosis, skewness, tail index,
and asymmetry index (\citealt{bird1993}), the analysis of the velocity
distribution shows no evidence of possible deviations from Gaussian
distribution. Moreover, the BCG is a good indicator of the mean
redshift of the cluster as  there is no evidence for a peculiar
velocity according to the Indicator test of \citet{gebhardt1991}.

\begin{figure}
%\centering
\resizebox{\hsize}{!}{\includegraphics{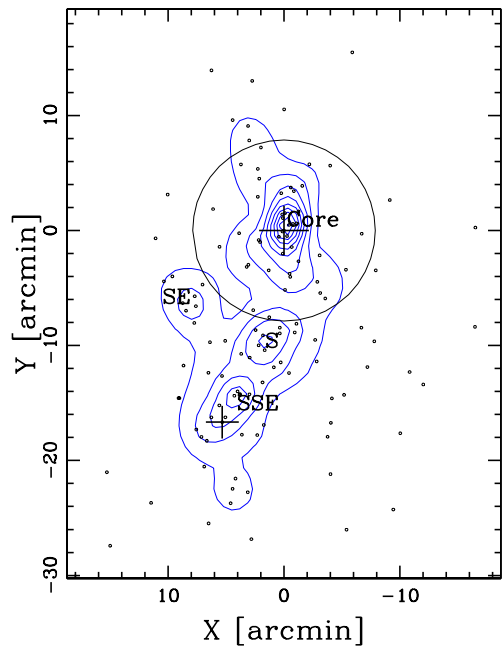}}
\caption{Spatial distribution of the 126 spectroscopic-determined
  cluster members on the sky and the relative isodensity contour map
  obtained with the 2D DEDICA method (black dots and blue
  contours). The two black crosses indicate the position of the BCG
  and LEDA~87445.  The plot is centered on the cluster center and
  circle contains the cluster within a radius of  0.5 Mpc $\sim$
  $1/3 \times R_{200}$. Labels refer to the 2D DEDICA peaks in the
  galaxy distribution shown in Tab.~\ref{tabdedica2dz}.  }
\label{figk2z}
\end{figure}

We analyzed the spatial distribution of the 126 spectroscopic member
galaxies  using the 2D adaptive-kernel method of Pisani et
al. (\citeyear{pisani1996}, hereafter 2D-DEDICA, see also Girardi et
al. \citeyear{girardi1996}).  Our results are shown in
Fig.~\ref{figk2z} and Tab.~\ref{tabdedica2dz}.  For each clump of galaxies
detected with a c.l. higher than 99\%,
Table~\ref{tabdedica2dz} contains the number of member galaxies, the
position of the 2D density peak, and its relative density $\rho$ with
respect to the densest peak.  Figure~\ref{figk2z} shows a peak in
the cluster core (Core) and other peaks in the south and southeast (S,
SSE, and SE). The BCG and LEDA~87445 belong to Core and SSE galaxy
groups, respectively. To verify that our results are independent of the
positioning of the spectroscopic masks, we performed the same analysis for the
497 non-member galaxies and found that non-members cluster in a different way.

\begin{table*}
        \caption[]{2D DEDICA substructure}
         \label{tabdedica2dz}
            $$
         \begin{array}{l r c c c}
         \hline
         \noalign{\smallskip}
         \hline
         \noalign{\smallskip}

\mathrm{Group} & N_{\rm gal} & \alpha({\rm J}2000),\,\delta({\rm J}2000)&\rho&\mathrm{<V>}\\
& & \mathrm{h:m:s,\degree:\arcmm:\arcs}&&\mathrm{km\ s^{-1}}\\

          \hline
           \noalign{\smallskip}

\mathrm{Core~(BCG)}   & 53&09\ 18\ 04.8,-12\ 05\ 20&1.00&15990\pm198\\ 
\mathrm{S}            & 20&09\ 18\ 12.3,-12\ 15\ 17&0.46&16279\pm189\\      
\mathrm{SSE~(LEDA)}   & 22&09\ 18\ 22.3,-12\ 20\ 05&0.51&16766\pm183\\      
\mathrm{SE}           &  9&09\ 18\ 38.6,-12\ 12\ 05&0.27&16663\pm232\\         

           \noalign{\smallskip}
           \hline
           \end{array}
           $$
\end{table*}

For each group, Table~\ref{tabdedica2dz} lists the mean velocity of
the member galaxies. The mean velocity of the galaxies in the SSE
group is $3\sigma$ higher than that of the galaxies in the Core group
($\Delta V_{\rm rf}\sim 850$ \ks at the cluster rest frame). The
velocity difference is of the same order as that between the two
brightest galaxies, the BCG and LEDA~87445. However, the BCG velocity
is significantly higher than the mean velocity of the Core group, with
a c.l. in the range 95-99\% according to the Indicator test.  The
velocity of LEDA~87445 is higher than that of the SSE group, with
a c.l. in the range 90-95\%.  These discrepancies could be the
consequence of a truly complex structure or contamination by
projected galaxies. To address the latter issue, we present below
analyses based on both positions and velocities.

\begin{figure}
\centering 
\includegraphics[width=8cm]{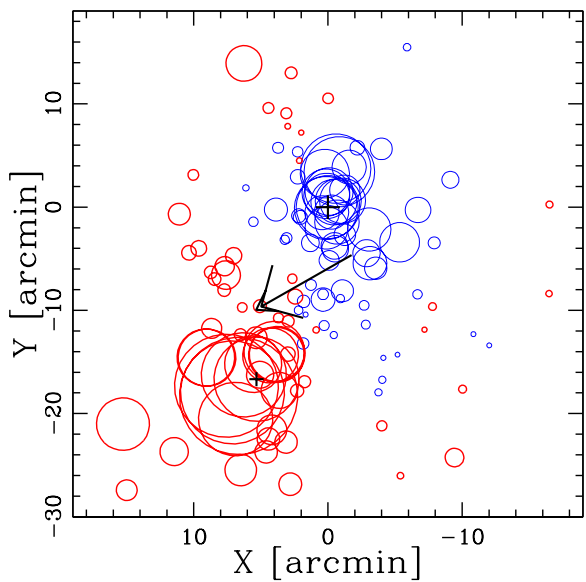}
%\includegraphics[width=8cm]{figdssegno5v.eps}
%\resizebox{\hsize}{!}{\includegraphics{figdssegno10v.eps}}
%\resizebox{\hsize}{!}{\includegraphics{figdssegno5v.eps}}
\caption
{Dressler \& Schectman bubble plot for the DSV test. Spatial
  distribution of the 126 cluster members, each indicated by a symbol:
  the larger the symbol, the greater the deviation $|\delta_{i,{\rm V}}|$ of
  the local mean velocity from the global mean velocity.  The blue
  thin and red thick circles indicate where the local mean velocity is smaller
  or larger than the global mean velocity.  The diagram is centered on the
  center of the cluster and the large and small crosses indicate the
  BCG and LEDA~87445.  The arrow indicates the direction of the
  velocity gradient, which points towards the region with high
  velocities.  }
\label{figds10v}
\end{figure}

As for the full 3D analysis, we looked for a correlation between
velocity and position information.  The presence of a velocity
gradient is quantified by a multiple linear regression fit to the
observed velocities with respect to the galaxy positions in the plane
of the sky (see also \citealt{denhartog1996}).  The position angle on
the celestial sphere is $PA=120_{-14}^{+20}$ degrees (measured
counterclockwise from the north), which means that high-velocity
galaxies are located in the southeastern region of the cluster (see
Fig.~\ref{figds10v}).  To assess the significance of this velocity
gradient, we performed 1000 Monte Carlo simulations of clusters by
randomly shuffling the velocities of the galaxies. For each
simulation, we determine the coefficient of multiple determination
($RC^2$, \citealt{nag1986}).  The significance of the velocity
gradient is the fraction of cases in which the $RC^2$ of the simulated
data is smaller than the observed $RC^2$.  In A780, the velocity
gradient is significant at the $98.5\%$ confidence level.

We have used the classical $\Delta$-test of Dressler \& Shectman
(\citeyear{dressler1988sub}, DS test).  For each $i$-th galaxy, the
deviation of the local mean velocity from the global velocity is
defined as $|\delta_{i}|$ with $\delta_{i}^2=[(N_{\rm
    nn}+1)/\sigma^2_{\rm V}]\times [(\left<V\right>_{\rm loc}
  -\left<V\right>)^2+(\sigma_{\rm V,loc}-\sigma_{\rm V})^2]$, where
the local mean velocity $\left<V\right>_{\rm loc}$ and velocity
dispersion $\sigma_{\rm V,loc}$ are calculated using the $i$-th galaxy
and its $N_{\rm{nn}}=10$ neighbors. For a cluster, the cumulative
deviation is given by the value of $\Delta$, which is the sum of the
$|\delta_i|$ values of the individual $N$ galaxies.  We have also used
the modified version that considers only the indicator of local mean
velocity, that is $\delta_{i,{\rm V}}=[(N_{\rm nn}+1)^{1/2}/\sigma_{\rm
    V}]\times (\left<V\right>_{\rm loc} -\left<V\right>)$ and
$\Delta$ is the sum of the
$|\delta_{i,{\rm V}}|$ values of the individual $N$ galaxies (DSV test,
\citealt{girardi2010}). As in the calculation
of the velocity gradient, the significance of the $\Delta$, i.e.  the
presence of substructure, is based on 1000 Monte Carlo simulated
clusters. In A780, the significance of the substructure is larger than
the 99.9\% c.l. according to both DS and DSV tests.  In
Fig.~\ref{figds10v} we show the Dressler \& Schectman bubble-plot
resulting from the indicator of the DSV test, $|\delta_{i,{\rm V}}|$.  This
plot shows very clearly how the galaxy LEDA~87445 is surrounded by
galaxies that have higher velocities than the galaxies in the core of
the cluster.

\begin{figure}
\centering 
\resizebox{\hsize}{!}{\includegraphics{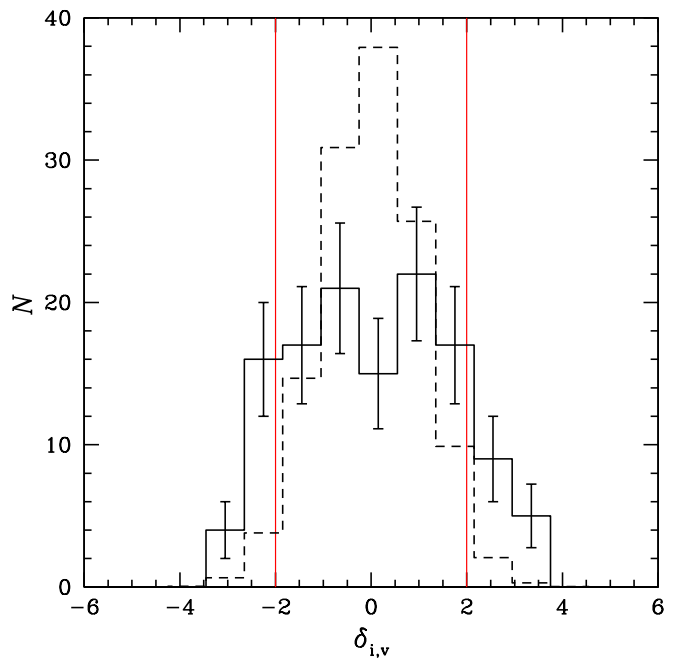}}
\caption
    {Distribution of $\delta_{i,{\rm V}}$ values of the deviation of the
      local mean velocity from the global velocity (according to the 
      DSV test, see text).  The solid line histogram indicates the observed
      galaxies. The dashed line histogram indicates the galaxies of
      the simulated clusters, normalized to the number of observed galaxies. The
      red vertical lines mark the $|\delta_{i,{\rm V}}|>2$
      regions where we expect to find substructure members.}
\label{figdeltai}
\end{figure}

As a first attempt to detect substructure members, we resorted to the
technique developed by Biviano et al. (\citeyear{biviano2002}). We
compared the distribution of $\delta_{i,{\rm V}}$-values of the real
galaxies with the distribution of $\delta_{i,{\rm V}}$-values of the
galaxies of all 1000 Monte Carlo simulated clusters
(Fig.~\ref{figdeltai}). According to the Kolmogorov-Smirnov test, the
two distributions differ at the $95\%$ confidence level.  The
distribution of the values of the real galaxies shows a tail at low
$\delta_{i,{\rm V}}$ values and a tail at high $\delta_{i,{\rm V}}$ values. When
considering galaxies with $|\delta_{i,{\rm V}}|>2$, the 15 galaxies in the
low tail are projected onto the core region of the cluster and the 14
galaxies in the high tail are projected onto the LEDA~87445 region.

\begin{figure}
\centering
\resizebox{\hsize}{!}{\includegraphics{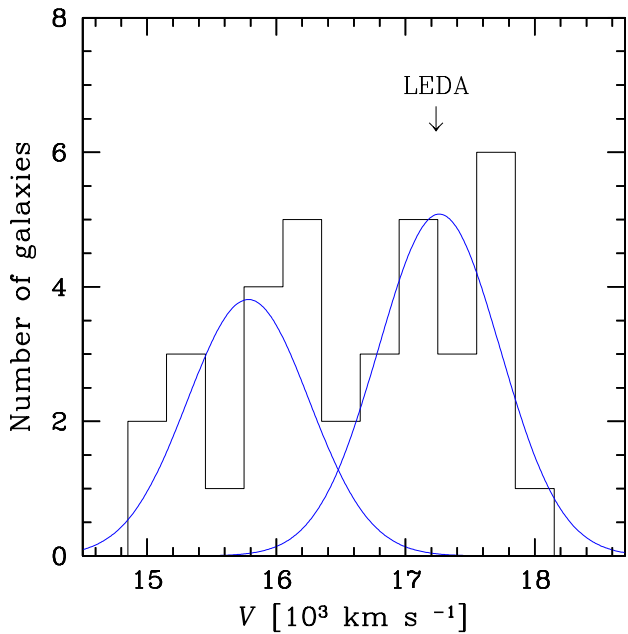}}
\caption
{Velocity distribution of all 35 galaxies within 0.5 Mpc of
  LEDA~87445.  The blue curves show the two Gaussian curves resulting from
  the 1D KMM analysis for the two detected groups LVG05 and HV05.  }
\label{fighistoleda05mpc}
\end{figure}

In order to better determine the members of the substructures, in
particular the group around LEDA~87445, we also analyzed our
data using three methods.  In  a first step, we analyzed the sample
of 35 galaxies within 0.5 \mpc of LEDA~87445 using the Kaye's
Mixture Model (KMM).  This method, as implemented by
\citet{ashman1994}, is useful for finding a possible group partitioning of
the velocity distribution. The KMM algorithm fits a user-specified
number of Gaussian distributions to a data set and evaluates the
improvement of this fit over a single Gaussian according to the
likelihood ratio test statistic. The algorithm also makes an
assignment of galaxies into groups.  When the KMM method is applied to
the sample of 35 galaxies, we find that a bimodal distribution is
better than a unimodal one at the $\sim 95\%$ confidence level.
We obtain a low-velocity group (LVG05) of 15 galaxies
with $\left<V\right>=15745$ \kss and a high-velocity group (HVG05)
of 20 galaxies with $\left<V\right>=17273$ \ks (see
Fig.~\ref{fighistoleda05mpc}).  The high velocity group contains
LEDA~87445.

\begin{figure}
\centering
\resizebox{\hsize}{!}{\includegraphics{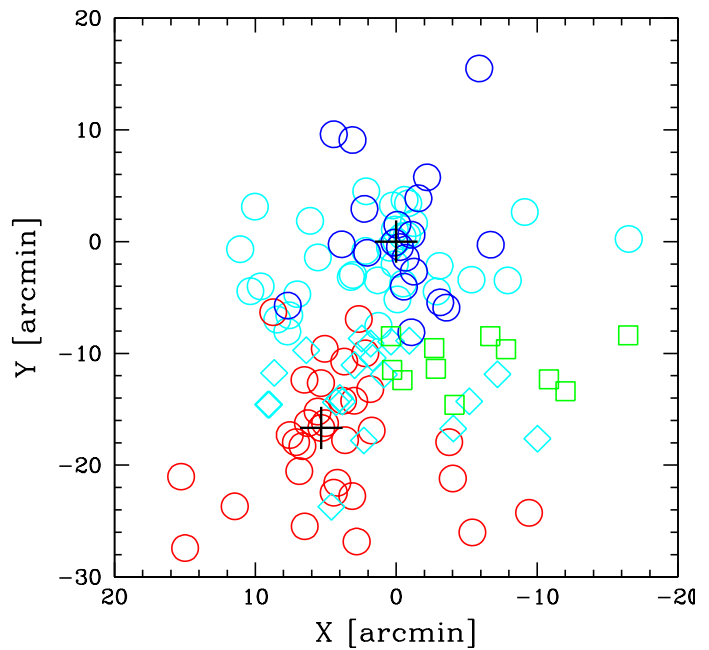}}
\caption{Spatial distribution of the 126 cluster members on the sky, with
  the groups discovered with the 3D DEDICA method indicated by
  different symbols.  From low to high velocities: 3DLVG (blue
  circles), 3DMVG1 (cyan rotated squares), 3DMVG2-BCG (cyan circles),
  3DMVG3 (green square), and 3DHVG-LEDA (red circles).  Crosses
  indicate the BCG and LEDA~87445. }
\label{figded3dpis}
\end{figure}

We also used the 3D DEDICA method (\citealt{pisani1996};
\citealt{bardelli1998}).  Table~\ref{tabdedica3dpis} lists the
properties of the galaxy peaks that are detected with a probability
higher than the $>99\%$ c.l. and have a relative density $\rho\ge
0.3$: the number of member galaxies, the positions of the peaks in
velocity and spatial coordinates, the relative
densities. Figure~\ref{figded3dpis} shows the positions of the
galaxies assigned to the detected groups.  The group with the lowest
velocity (3DLVG) is projected onto the core of the cluster. The
highest velocity group (3DHVG) is the one projected onto the
south-southeast region and contains LEDA~87445.  Of the three
intermediate velocity groups, 3DMVG2 is projected onto the core of the
cluster and contains the BCG.  Since it is known that the
multidimensional application of the DEDICA algorithm may lead to
spurious substructure (\citealt{bardelli1998}), we also applied the
alternative version of \citet{balestra2016} based on the rule of thumb
for the kernel size given by \citet{silverman1986}. The goal is to
detect the major substructures at the cost of losing some real minor
substructures. The alternative procedure is successful in recovering a
main system (96 galaxies) and the 3DSHVG group (30 galaxies). The
3DSHVG group contains LEDA~87445 and is populated by the same galaxies
as the 3DHVG group, with the exception of two galaxies. This suggests
that the detection of a galaxy group related to LEDA~87445 is quite
robust.

     \begin{table}
        \caption[]{3D DEDICA substructure}
         \label{tabdedica3dpis}
            $$
         \begin{array}{l r r c c }
            \hline
            \noalign{\smallskip}
            \hline
            \noalign{\smallskip}

\mathrm{Group} & N_{\rm gal} & V_{\rm peak}& \alpha({\rm J}2000),\,\delta({\rm J}2000)&\rho\\
& & {\rm km\,s^{-1}}&\mathrm{h:m:s,\degree:\arcmm:\arcs}&\\

         \hline
         \noalign{\smallskip}

\mathrm{3DMVG2(BCG)}  & 34&15812&09\ 18\ 04.6,-12\ 05\ 12&1.00\\
\mathrm{3DHVG(LEDA)}  & 32&16932&09\ 18\ 25.3,-12\ 21\ 06&0.30\\
\mathrm{3DMVG1}       & 20&15761&09\ 18\ 08.4,-12\ 14\ 41&0.54\\
\mathrm{3DLVG}        & 20&15521&09\ 18\ 04.3,-12\ 06\ 55&0.86\\
\mathrm{3DMVG3}       & 11&16406&09\ 18\ 01.7,-12\ 15\ 44&0.33\\
              \noalign{\smallskip}
              \hline
         \end{array}
$$
\end{table}

\begin{figure}[!ht]
\centering 
\includegraphics[width=9cm]{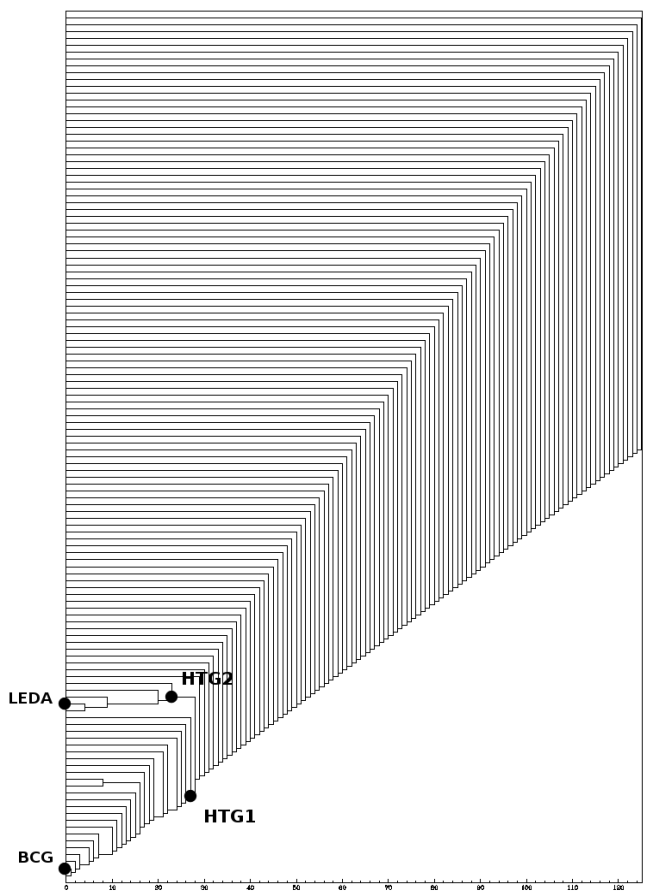}
\caption{Dendrogram obtained by the Htree algorithm.  The x-axis
  is the binding energy, here in arbitrary units with the lowest
  negative energy levels on the left. The labels indicate prominent
  galaxies and structures.}
\label{figAMMgerbal}
\end{figure}

As a final approach, we applied the method described by
\citet{serna1996}, which is referred to in the literature as the
Serna-Gerbal or H-tree method (e.g.,
\citealt{durret2009},\citealt{girardi2011},\citealt{guennou2014}).
The H-tree method uses hierarchical cluster analysis to determine the
relationship between galaxies based on their relative binding
energy. The ignorance of the mass associated with each galaxy is
overcome by assuming the typical mass-to-light of galaxy clusters for
each galaxy halo.  Figure~\ref{figAMMgerbal} shows the resulting
dendrogram for the case $M/L_r$=200 \ml
(e.g., \citealt{popesso2005}), where the total energy appears
along the horizontal axis.  At the $\sim 30$ energy level (in normalized units),
the cluster splits into two groups, HTG1 and HTG2.  The BCG and its
close companion are located in the potential hole of the HTG1 group,
which is populated by 24 galaxies.  LEDA~87445 is located in the
potential hole of group HTG2, which is populated by five
galaxies. HTG1 and HTG2 probably correspond to the cores of the main
cluster and the LEDA~87445 group, respectively. Similar results are
found in the cases with $M/L_r$=150 and 250 \mll. Table~\ref{tableda}
summarizes the kinematic properties of the LEDA group obtained
by different methods.

\begin{table*}
        \caption[]{Kinematic properties of the group related to LEDA\,87445}
         \label{tableda}
            $$
         \begin{array}{l r c l l l }
            \hline
            \noalign{\smallskip}
            \hline
            \noalign{\smallskip}

\mathrm{Group}^{\mathrm{a}}& N_{\rm gal} &^{\mathrm{b}} \alpha({\rm J}2000),\,\delta({\rm J}2000)&\mathrm{<V>}&\sigma_{\rm V}&\mathrm{Method\ of \ detection}\\
& & \mathrm{h:m:s,\degree:\arcmm:\arcs} &\mathrm{km\ s^{-1}}&\mathrm{km\ s^{-1}}&\\
         \hline
         \noalign{\smallskip}

\mathrm{HVG05} &20&09\ 18\ 24.53,-12\ 20\ 52.2 &17273\pm97&419_{-52}^{+46} &\mathrm{1D-KMM\,within\,0.5\,Mpc}\\
\mathrm{3DHVG} &32&09\ 18\ 25.18,-12\ 23\ 25.9 &17140\pm92&509_{-56}^{+56} &\mathrm{3D-DEDICA}\\
\mathrm{3DSHVG}&30&09\ 18\ 25.22,-12\ 22\ 58.2 &17193\pm89&478_{-52}^{+54} &\mathrm{3D-DEDICA-Silvermann}\\
\mathrm{HTG2} &5&09\ 18\ 29.15,-12\ 21\ 52.1 &17285\pm293&499_{-41}^{+322} &\mathrm{Htree}\\

            \noalign{\smallskip}
            \hline
         \end{array}
$$
\begin{list}{}{}  
\item[$^{\mathrm{a}}$] For comparison, the center and velocity of LEDA\,87445 are
09\ 18\ 27.44-12\ 22\ 24.0 and $17236\pm 46$ km\ s$^{-1}$, respectively. 
\item[$^{\mathrm{b}}$] As centers, 
  we list the biweight mean values of   galaxy right ascensions and declinations.
\end{list}
         \end{table*}

\section{Discussion}
\label{discu}

\subsection{Cluster global  properties}
\label{discuglob}

First, we discuss our results on global properties of A780. Our
estimates for the cluster mean redshift,
$\left<z\right>=0.0545\pm0.0002$, and velocity dispersion,
$\sigma_{\rm V}=795_{-53}^{+44}$ \kss, agree within $1\sigma$ with our
previous estimates in \citet{degrandi2016}. With four times as many
cluster members, the uncertainties are smaller by a
factor of two. Using our estimate of $\sigma_{\rm V}$ and the estimate
of the global X-ray temperature in the literature ($kT_{\rm X}\sim
3-3.5$ keV), we can calculate the ratio of the energy per unit mass of
galaxies and ICM, $\beta_{\rm spec}=0.97_{-0.21}^{+0.26}$, where
$\beta_{\rm spec}=\sigma_{\rm V}^2/(kT_{\rm X}/\mu m_{\rm p})$ with
$\mu=0.58$ the average molecular weight and $m_{\rm p}$ the proton
mass. Since $\beta_{\rm spec}=1$ is expected when assuming
density-energy equipartition between galaxies and ICM, our result
means that A780 is probably not far from the dynamical
equilibrium. This indication of large-scale relaxation is consistent
with evidence reported in the literature, such as the presence of a
cool core and the smoothness of the X-ray morphology of A780 within 1
Mpc (see Sect.~\ref{intro}).

The velocity of the BCG is very close to the mean velocity of the
galaxy cluster and we have found that there is no peculiarity in terms
of the velocity distribution of the cluster galaxies. For the BCG, we
calculate $|V_{\rm rf}/\sigma_{\rm V}|=0.08$, which places A780 in the
low tail of the $|V_{\rm rf}/\sigma_{\rm V}|$ distribution, as
obtained by \citet{lauer2014} for 178 clusters. The BCG
spectrum shows a few emission lines (see Fig.~\ref{figneichel}) and
the BCG color is bluer than that of the red sequence (see
Fig.~\ref{figcm}), which is consistent with what is expected for BCGs
in cool core clusters, especially for BCGs with radio emission, as
known in the literature (e.g., \citealt{crawford1995}).  Thus, the BCG
also exhibits the typical features of BCGs in relaxed systems.

Our estimate of the dynamical mass, $M_{200}=5.4\pm1.5$ \mquaa, is in
agreement with the mass obtained from the gravitational lensing,
$M_{200,\rm{GL}}=3.7_{-1.4}^{+2.0}$ \mqua
(\citealt{okabe2016}). Taking into account the large uncertainties,
there is also a $2\sigma$ agreement with the mass estimate from X-ray
data, $M_{200,\rm{X}}=2.84_{-0.27}^{+0.30}$ \mqua
(\citealt{zhu2021}). However, our nominal value of the dynamical mass
is a factor of two higher than the value of the mass from X-rays, and
we prefer to discuss it a little more.  Mass estimates of galaxy
clusters are subject to various systematic uncertainties and biases,
depending on the approach taken (see \citealt{pratt2019} for a
review).  This can justify strong differences when comparing mass
estimates for a single cluster. Indeed, when samples of many clusters
with high quality data are examined, generally better agreement is
found (\citealt{ettori2019}).

Although our study is not dedicated to the study of the galaxy
population, our data show that the red sequence is visible at least
down to $r\sim$ 21 mag, about five magnitudes below  the
value of $M^*$ in the luminosity function (see Fig.~\ref{figcm}).
This is consistent with previous studies that have looked at the
spectroscopically
determined luminosity function in clusters down to similarly deep
magnitudes.  These studies show that the red
sequence in local clusters is already established down to galaxies with faint
magnitudes, albeit with different levels of definition
(\citealt{rines2008}; \citealt{aguerri2020} and refs. therein). We
refer the reader to the review by \citet{boselli2014} for a full discussion
of the origin of the faint-end of the red sequence in high-density
environments.

\subsection{Cluster substructure}
\label{discusub}

Our study is designed to provide new insights into the small-scale
structure of A780, especially in the LEDA~87445 region.  We present
solid evidence for the existence of a group of galaxies around
LEDA~87445.  The first evidence comes from our 2D analysis (see
Fig.~\ref{figk2z}).  It is useful to compare our present results with
those based on photometric likely members in our previous study, which
uniformly cover the entire cluster field (see Fig.~A.3 in
\citealt{degrandi2016}). Both the old and new maps of the galaxy
distribution show the presence of a substructure related to
LEDA~87445. This suggests that photometric likely members can be a
good alternative to study the internal structure of the cluster, at
least if the bright galaxies located in the red sequence are
considered (e.g., \citealt{lubin2000}).  The final evidence comes from
the the DS and DSV tests, which find a very significant correlation
between galaxy positions and velocities, confirming the presence of
high-velocity galaxies around LEDA~87445 (see Fig.~\ref{figds10v}).

Our spectroscopic catalog is extensive enough to make an attempt to
identify the members of the LEDA~87445 group and to obtain a first
estimate of its kinematic properties (see Tab.~\ref{tableda}).  For
each method, we find good agreement between the mean
velocity of the group and the velocity of LEDA~87445, and between the center of
the group and the position of LEDA~87445. LEDA~87445 is
characterized by a higher velocity with respect to the cluster mean 
velocity, with a velocity difference of $V_{\rm rf} \sim$ +870 \kss.
Depending on the detection method, the velocity difference between
the group and the cluster ranges from $V_{\rm
  rf}=$+785\ks to +920 \kss.  LEDA~87445 lies $\sim$ 1.1 Mpc south of
the cluster center.  Our estimates of the group center are around
LEDA~87445, at distances in the range of 40-100 kpc.  We thus 
find no evidence of a systematic shift between the galaxies of the group
and LEDA~87445 itself, which can be considered a good tracer of the
group.

In \citet{degrandi2016}, we propose that the LEDA~87445 group is
infalling and moving in a circular orbit with a velocity of
$510_{-340}^{+430}$ \ks relative to the main cluster, as seen from
X-ray features. This velocity estimate is subject to large
uncertainties and no difference can be detected with the $V_{\rm rf}$
estimated here. In \citet{degrandi2016}, we also propose that the tail
of the LEDA~87445 group is due to ram pressure.  According to new
hydrodynamical simulations, \citet{sheardown2019} claim that the
X-ray features are more consistent with the scenario of a slingshot
tail rather than a tail caused by ram pressure. In this scenario, the
group has already passed the cluster center with a large impact
parameter and is now near the apocenter, where it re-enters the
cluster (see their Fig.~1).  However, \citet{sheardown2019} do not give
an estimate of the velocity of the LEDA~87445 group.

Table~\ref{tableda} also reports the estimates for the velocity
dispersion of group galaxies. We emphasize that the estimate of the
dispersion is less robust than the estimate of the mean. In
particular, possible contamination by non-member galaxies is likely to
lead an overestimation of the true value of the velocity dispersion.
The nominal values of the velocity dispersion range from $\sigma_{\rm
  V} \sim$420 to $\sim$500 \kss, that is $kT_{\rm X} \sim 1.1$-1.5
keV assuming $\beta_{\rm spec}=1$. This temperature estimate is in
agreement with X-ray data for the gas along the wake ($kT_{\rm X} \sim
1.3$, \citealt{degrandi2016}).  Our estimate of the group mass, based
on the value of $\sigma_{\rm V}$ and the relation of
\citet{munari2013}, is $M_{200}=0.8$-1.3 \mquaa, while the mass
estimated by \citet{degrandi2016} using the $M_{500}$-$kT_{\rm X}$
relation of \citet{sun2009} was found to be $M_{\rm 500,X}=0.2$-0.5
\mquaa, i.e. $M_{\rm 200,X}=0.3$-0.7 \mquaa.  In the case of the
LEDA~87445 group, the main source of the uncertainty in the dynamical
mass is probably related to the selection of the group galaxies.  On
the other hand, the gas content might suffer more from the interaction
with the main system due to its collisional nature, and therefore the
X-ray mass is also quite uncertain.  Taking our nominal values of the
mass for the LEDA~87445 group and the parent cluster, we can estimate
that their mass ratio is $\sim 1:5$.

We also obtain new information about the cluster core.  The galaxy
distribution in the cluster core is characterized by some strong
peculiarities.  The phase-space diagram of the cluster galaxies shows
that the BCG velocity is close to the cluster mean velocity (set to
zero in Fig.~\ref{figvd}), but in the inner regions, within $\sim 0.5$
Mpc, galaxies with velocities larger than that of the BCG are very
few.  This asymmetry is real with a probability of  $99.1\%$, as
determined by applying a 2D Kolmogorov-Smirnov test to compare the
distributions of galaxies with positive and negative velocities within
0.5 Mpc. The mean velocity of the 44 galaxies within 0.5 Mpc is
  $\left<V_{\rm{rf}}\right>=-259 \pm 107$ \kss, which differs from the
  expected zero value at $99.3\%$, in agreement with the above test.
Note for comparison that the distribution of galaxies in the phase
space outside 0.5 Mpc does not show significant asymmetry between
positive and negative velocities, despite the presence of the
LEDA~87445 group.  Moreover, according to the DSV test, the cluster
core is characterized by galaxies with peculiar low velocity (see
Fig.~\ref{figds10v}) and the BCG has a higher peculiar velocity with
respect to the galaxy peak detected by our 2D DEDICA analysis in the
core.  Another indication of a peculiarity of the mass distribution in
the cluster core comes from the analysis of the weak lensing data,
which are better described at the innermost radius by a two-component
model (\citealt{okabe2016}).

A possible interpretation of the above features in the cluster
core is that there is a low-velocity group, still far from being
virialized within the cluster, which is projected onto the cluster
core.
Indeed, using the 3D DEDICA method, we can
detect a group of low-velocity galaxies in the core (3DLVG) whose peak
velocity is lower than the mean velocity of the cluster, $V_{\rm
  {rf}}\sim -750$ \kss, and the galaxy content is $\sim 10\%$ of
  the whole system (see Tab.~\ref{tabdedica3dpis} and Fig.~\ref{figded3dpis}).

To check the above ``group hypothesis'', we made a simple Monte
  Carlo simulation to reproduce a combined system of cluster plus a group
  with the same center.  For a given mass $M_{200}$, we use the NFW
  model with the appropriate concentration parameter
  (\citealt{navarro1997}; \citealt{dolag2004}) to simulate the 2D
  galaxy distribution of galaxies out 2$R_{200}$ (e.g.,
  \citealt{biviano2003}; \citealt{rines2013}). The velocities are
  distributed according to a Gaussian.  To mimic the cluster we use
  the mass and velocity dispersion of the whole A780 ($M_{200}=5.4$
  \mquaa, $\sigma_{\rm V}=795$ \kss) and distribute the particles in
  such a way to have 10000 particles within $R_{200}=1.65$ \hh, with
  zero mean velocity. The group is ten times less massive
  ($M_{200}=0.54$ \mquaa, $\sigma_{\rm V}=369$ \kss), with 1000
  particles within $R_{200}=0.767$ \hh, and mean velocity equal to
  $-750$ \kss.  Out of this combined system we randomly extract 1000
  times 44 particles within $R=0.5$ \h from the center to obtain 1000
  values of mean velocity, $MV_{\rm{rf,sim}}$.  The value of
  $\left<MV_{\rm{rf,sim}}\right>$ is $-119$ \kss, which differs at
  $\sim 90\%$ from the observed value $\left<V_{\rm{rf}}\right>=-259
  \pm 107$ \kss.  In summary, the presence of a group projected onto
  the cluster center can strongly reduce the significance of the
  observed asymmetry in the velocity distribution.  We stress that our
  model is very simple. For instance, we assume that the particle
  number-density rescale with the mass of the system, while the very
  luminous BCG in A780 may indicate a different rescaling. Very
  luminous BCGs might be formed at the expenses of the
  merger/cannibalization of surrounding galaxies with a subsequent
  change of the luminosity function (e.g., see
  \citealt{zarattini2015}). This would make the effect of the group
  more important.

  Using N-body simulations, \citet{vijayaraghavan2015} show that the core of
  a group survives for many Gyr after a head-on merger with a cluster.
  On the other hand, the presence of a group--cluster interaction
  would raise a new issue since cool cores in clusters can be
  resilient for off-axis mergers but not for head-on ones
  (\citealt{valdarnini2021} and refs. therein). Therefore we suggest
  that the low-velocity group, if confirmed, might be on its first
  infall onto the cluster.

Finally, we discuss our results in terms of the central engine of the
Hydra~A AGN. The shock detected in the \emph{Chandra} X-ray images,
which is modeled as an ellipse with a size of $\sim 400$ kpc, shows a
displacement of the center with respect to the AGN, and this
displacement was caused by bulk flows or sloshing in the ICM according
to \citet{simionescu2009}.  According to the authors, these gas
motions should be associated with velocities of the order of 600 \kss
and could be related to a small merger.  We detect a galaxy group
  related to LEDA~87445 group with a comparably high relative velocity
  and the simulation of \citet{sheardown2019} explains the
  sloshing of the ICM at the cluster center, in particular when the
  merging group is at its apocenter (see also the third column of
  Fig.~7 in \citealt{sheardown2018}).

\section{Summary and conclusions}
\label{summa}

Our previous study (Appendix~A of \citealt{degrandi2016}) was based on
the largest available homogeneous sample of spectroscopic data at the
time, with 41 galaxies (\citealt{smith2004}), from which we selected
33 member galaxies.  To surpass our earlier study, we have acquired
new spectroscopic data at TNG and VLT, giving us a redshift catalog of 623
galaxies.  Our new spectroscopic data are mainly focused on
the south-southeast region of the cluster, i.e.  the region of the
bright X-ray tail.  We have paid special attention to combining data
from different sources in order to obtain a homogeneous redshift
catalog.  Our membership selection procedure leads to a catalog of 126
member galaxies, which extends the data sample analyzed in our
previous study by a factor four.  In the following, we list our main
results.

\begin{enumerate}
  
\item We derive the mean redshift of the cluster
  $\left<z\right>=0.0545\pm0.0002$ and the line-of-sight velocity
  dispersion $\sigma_{\rm V}=795_{-53}^{+44}$ \kss. We estimate the dynamical
  mass $M_{200}=5.4\pm1.5$ \mqua within
  $R_{200}=1.65\pm0.10$ Mpc.  We find no evidence for a deviation in
  the density-energy equipartition between galaxies and ICM or for a
  peculiarity of the BCG, consistent with the cluster being relaxed
  on a  large scale.
  
\item Using four different techniques, we detect the presence of a
  group related to LEDA~87445 and list the fiducial members. For each
  technique, we calculate the velocity and center  of the
  group. The difference between these estimates and the velocity and
  position of LEDA~87445 is less than 90 \ks and 0.1 Mpc,
  respectively. Therefore, LEDA~87445 is an excellent tracer of the
  galaxy group.  Our estimate of the velocity dispersion of the group
  is $\sim$ 420 -- 500 \kss, leading to a group mass 
  $M_{200}=0.8$--1.3 \mquaa.

\item The parameters of the collision can be summarized with a
  relative line-of-sight velocity of $V_{\rm rf}\sim +870$ \kss, a
  projected distance $D=1.1$ Mpc, and a mass ratio of 1:5 between the
  group and the cluster.

\item We present evidence of an asymmetry in the velocity
  distribution of galaxies in the inner cluster region. This asymmetry
  might be related to the presence of a small low-velocity group, as
  suggested by the detection of a substructure projected onto the
  cluster center at $V_{\rm rf}=-750$ \kss.

\end{enumerate}
  
  We conclude that A780, although dynamically relaxed at first sight,
  contains minor substructures. As for the LEDA~87445 group,
    optical and X-ray data result in a consistent scenario and
    previous studies have addressed its impact on the energetics of
    the core region.  To better understand the observed velocity
    asymmetry in the cluster core, we stress the need for additional
    redshifts of galaxies in that region.

  \begin{acknowledgements}
    We thank the referee for her/his useful and constructive comments.
    M.G.  acknowledges the support from the grant MIUR PRIN 2017
    WSCC32 ``Zooming into dark matter and proto-galaxies with massive
    lensing clusters''.  This publication is based on observations
    made on the island of La Palma with the Italian Telescopio
    Nazionale Galileo (TNG), which is operated by the Fundaci\'on
    Galileo Galilei -- INAF (Istituto Nazionale di Astrofisica) and is
    located in the Spanish Observatorio of the Roque de Los Muchachos
    of the Instituto de Astrof\'isica de Canarias. This publication is
    based on observations collected at the European Southern
    Observatory under ESO programme 098.A-0807(A).  This publication
    uses public data from the CFHT Science Archive, which contains
    data provided by the Canada-France-Hawaii Telescope.
\end{acknowledgements}

\bibliographystyle{aa}
\bibliography{biblio}

\end{document}